\documentclass[aps,PRB,reprint,superscriptaddress,preprintnumbers]{revtex4-2}
\usepackage[T1]{fontenc}
\usepackage{times}
\usepackage{graphicx,color}
\usepackage{amsfonts,amsmath,amssymb,amsbsy}
\usepackage[colorlinks=true, linkcolor=blue, citecolor=blue, urlcolor=blue]{hyperref}
\usepackage{float}
\usepackage{lettrine}
\let\bm=\boldsymbol

\begin{document}

\title{Universal quantum computing based on magnetic domain wall qubits}

\author{Shuang Li}
\affiliation{School of Science and Engineering, The Chinese University of Hong Kong, Shenzhen, Guangdong 518172, China}

\author{Xichao Zhang}
\affiliation{Department of Applied Physics, Waseda University, Okubo, Shinjuku-ku, Tokyo 169-8555, Japan}

\author{Motohiko Ezawa}
\email[Email:~]{ezawa@ap.t.u-tokyo.ac.jp}
\affiliation{Department of Applied Physics, The University of Tokyo, 7-3-1 Hongo, Tokyo 113-8656, Japan}

\author{Yan Zhou}
\email[Email:~]{zhouyan@cuhk.edu.cn}
\affiliation{School of Science and Engineering, The Chinese University of Hong Kong, Shenzhen, Guangdong 518172, China}

\clearpage

\begin{abstract}
Quantum computers allow to solve efficiently certain problems that are intractable for classical computers.
For the realization of a quantum computer, a qubit design as the basic building block is a nontrivial starting point.
We propose the utilization of nanoscale magnetic domain walls, which are stabilized by achiral energy, as the building blocks for a universal quantum computer made of ferromagnetic racetracks.
In contrast to the domain walls stabilized by conventional Dzyaloshinskii-Moriya interactions, these achiral domain walls are bistable and show two degenerate chirality forms. 
When the domain wall is extremely small, it can be viewed as a quantum mechanical object and the two degenerate chiralities of the domain walls can be used to encode the qubit states $\lvert 0 \rangle$ and $\lvert 1 \rangle$.
We show that the single-qubit quantum gates are regulated by magnetic and electric fields, while the Ising exchange coupling facilitates the two-qubit gates. 
The integration of these quantum gates allows for a universal quantum computation. 
Our findings demonstrate a promising approach for achieving quantum computing through spin textures that exist in ferromagnetic materials.

\end{abstract}

\date{\today}

\maketitle

\clearpage

\section{introduction}
A next generation computer will be built on a hybrid system including both quantum and classical technologies~\cite{awschalom2021development}.
Quantum computing has been proven to be more efficient when solving a specific class of problems compared to classical digital computers~\cite{shor1994algorithms,ladd2010quantum}.
At present, the technologies for creating and manipulating qubits are implemented in a number of different platforms, including trapped ions~\cite{cirac,monroe}, nuclear spins~\cite{kane,vandersypen,nuclear}, superconducting circuits~\cite{nakamura,arute}, quantum dots~\cite{dots}, spin textures~\cite{xia2022,christina2021,psaroudaki2022skyrmion,xia2023,zou2022domain}, and nano-electromechanical systems~\cite{nanotube,nems}.
Among the many and varied proposals for constructing quantum computers, spin textures, such as meron~\cite{xia2022} and skyrmion~\cite{christina2021,psaroudaki2022skyrmion,xia2023}, occupy a special place because they are topologically protected and thus manipulated robustly at a low temperature.

A qubit encoded in a mobile domain wall, one of another type of spin textures, on a ferrimagnetic racetrack has been proposed very recently~\cite{zou2022domain}. 
Two topologically distinct states having opposite chirality on ferrimagnetic racetracks stand for two qubit states.
In this work, the Dzyaloshinskii-Moriya interaction (DMI), which promotes the formation of the N\'eel-type domain walls with a specific chirality~\cite{emori,jacob,hrabec}, acts as an effective magnetic field along the racetrack.
On the other hand, the quantum dynamics of ferromagnetic domain walls in one-dimensional ferromagnetic spin chains and the tunneling effects have been studied~\cite{shibata,freire,Galkina}, where two domain walls of opposite chirality are equally preferred in the system.
Especially, the energy competition, including the Heisenberg ferromagnetic interaction, a longitudinal and a transverse anisotropies, is considered in the system, while excluding the DMI. 
The quantum behavior of the domain wall is the same as that of the free particle~\cite{shibata}.
The quantum properties could be potential useful for quantum computing.

Domain walls can be manipulated and moved by various ways, such as magnetic fields~\cite{beach,Schryer,Mougin,luo2020current}, electric currents~\cite{Thiaville,emori,yamaguchi,Martinez2014current}, and thermal effects~\cite{jwj,wangxs,Martinez2007thermal,Torrejon}.
This mobility can be harnessed for a multitude of applications owing to their exceptional characteristics, such as stability and resistance to perturbations.
An essential application of magnetic domain walls lies in magnetic memory devices, where they play a crucial role in the storage and retrieval of data~\cite{parkin2015memory,parkin}. 
Furthermore, there are proposals to utilize magnetic domain walls in logic devices to execute conventional logic computations~\cite{logic,omari2019toward,luo2020current}.
Additionally, magnetic domain walls hold potential in unconventional neuromorphic computing~\cite{ababei,cui2020maximized,siddiqui2019magnetic}, which facilitates the parallel computing and resolves the limit of Moore’s law.
Quantum computation is another promising methodology for breaking the limit of Moore’s law.

In quantum circuit model of computation, a quantum gate is a basic quantum circuit operating on a small number of qubits. 
Quantum gates are the basic building blocks of quantum circuits.
With a minimal set of quantum gates that enable the execution of any quantum algorithm, universal quantum computing is possible.
To execute any quantum computation, one or two qubit gates are sufficient.
The Solovay-Kitaev theorem supports the notion that the phase-shift gate, the Hadamard gate, and the controlled-NOT (CNOT) gate are sufficient for universal quantum computation~\cite{deutsch1985quantum,dawson2005solovay}. 
Instead, it is enough to construct the phase-shift gate, the X rotation gate, and the Ising gate for universal quantum computation.

Quantum effect is considered when the size of domain wall is of the order of nanometers~\cite{Tatara1994}.
In this work, we propose a universal quantum computer based on the domain wall chirality by constructing a set of universal quantum gates. 
The left-handed and right-handed chiralities are assigned to be the quantum states $\lvert 0 \rangle$ and $\lvert 1 \rangle$.
The NOT gate is achieved by the reversible transformations between the qubit states $\lvert 0 \rangle$ and $\lvert 1 \rangle$ when applying the same magnetic field.
The phase-shift gate is constructed, where we change the phases of the qubits states $\lvert 0 \rangle$ and $\lvert 1 \rangle$ by controlling the electric field.
The Ising gate and the CNOT gate are also discussed, which are two-qubit gates.
Universal quantum computation is possible with the use of these operations.
The coherence time of the domain wall qubit is estimated in the microsecond region by calculating with the material parameters.
Domain wall qubits offer two distinct benefits when it comes to their practical application. 
First, the construction of domain walls on a nanoscale, as small as ten nanometers, can be accomplished~\cite{hehn1996,PAULUS2001180}. 
Additionally, the spin rotation of the domain wall in the transverse direction are consistent, which makes it convenient for measurement purposes.

\section{methods}
In this work, we consider a thin ferromagnetic racetrack with a length of 1000 nm, a width of 32 nm, and a thickness of 1 nm.
The bistable N\'eel-type domain walls can exist in ferromagnetic materials such as Co~\cite{Boehm2017} and CoFeB~\cite{Devolder2016}, which are sandwiched by heavy metals to eliminate net DMI due to the symmetric ferromagnetic interfaces.
We perform the micromagnetic simulations by using the GPU-accelerated micromagnetic simulator MUMAX3~\cite{mumax}.
The magnetization dynamics of the domain walls are controlled by solving the Landau–Lifshitz–Gilbert (LLG) equation without consideration of the thermal fluctuations,
\begin{equation}
\frac{d \bm{M}}{dt} = -\gamma \bm{M} \times \bm{h}_{\rm eff} + \alpha (\bm{M} \times \frac{d \bm{M}}{dt}),
\end{equation}
where $\bm{M}=M_S(\sin\theta\cos\phi, \sin\theta\sin\phi, \cos\theta)$ is the magnetization with the saturation magnetization $M_S$, $\gamma$ is the absolute gyromagnetic ratio, and $\alpha$ is the Gilbert damping parameter.
$\bm{h}_{\rm eff} = - \frac{1}{\mu_0 M_S} \frac{\partial \mathcal{E}}{\partial \bm{M}}$ is the effective field, where $\mu_0$ and $\mathcal{E}$ are the vacuum permeability constant and the total energy density, respectively.
The energy terms considered in the simulations include the exchange energy, the perpendicular magnetic anisotropy energy, and the demagnetization energy, where the demagnetization effect results in the magnetic shape anisotropy.
Correspondingly, the total energy $H$, a volume integral of the energy density $\mathcal{E}$, is
\begin{equation}
H =\int dV [A(\nabla\bm{M})^2-KM_z^2-\frac{\mu_0}{2}\bm{M}\cdot{\bm{H}_d}-\bm{M}\cdot{\bm{B}}],
\label{H1}
\end{equation}
where $A$, $K$, $\bm{H}_d$, and $\bm{B}$ denote the ferromagnetic exchange, perpendicular magnetic anisotropy constant, demagnetization field, and external magnetic field, respectively.
$M_z$ is the out-of-plane component of $\bm{M}$.
In all simulations, the ferromagnetic racetrack is discretized into 2 $\rm nm$ $\times$ 2 $\rm nm$ $\times$ 1 $\rm nm$ cells.
Following parameters are considered~\cite{Devolder2016}: the saturation magnetization $M_S = 1\times 10^6$ $\rm A m^{-1}$, the exchange constant $A = 3\times 10^{-11}$ $\rm J m^{-1}$, the out-of-plane anisotropy $K = 0.7\times 10^6$ $\rm J m^{-3}$, and the Gilbert damping parameter $\alpha = 0.1$.

\section{degenerate energies in the absence of magnetic field}
\label{degenerate}
\begin{figure}
\includegraphics{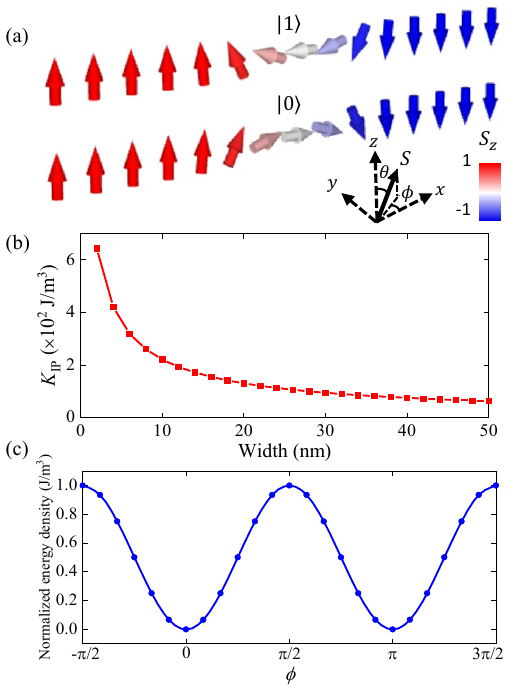}
\caption{(a) Illustration of N\'eel-type domain walls with different chiralities representing qubit states $\lvert 0 \rangle$ and $\lvert 1 \rangle$. The arrow denotes the spin direction and the color bar shows the out-of-plane spin components. 
(b) The effective in-plane anisotropy constant $K_{\rm IP}$ as a function of the width of the racetrack.
(c) The degenerate energies with different chiralities.
The in-plane anisotropy constant $K_{\rm IP}=1.4\times10^2$ $\rm J m^{-3}$ is used here.}
\label{1}
\end{figure}

The N\'eel-type domain walls that are stabilized by achiral energy terms instead of the usual DMI are bistable, with the two possible chiral forms being degenerate, as shown in Fig.~\ref{1}(a).
$\phi$ denotes the angle between the spin direction at the center of the domain wall and the $+x$ coordinate.
We demonstrate analytically that the in-plane magnetic shape anisotropy arising from the demagnetization interaction contributes to the degenerate energies.
To analyze the bistable states in the system, we assume that $\bm{M}$ is a function of $x$ alone and there is no magnetic field. The total energy Eq.~(\ref{H1}) of the domain wall system thus can be rewritten in polar coordinate system, considering the demagnetization energy in terms of the shape anisotropy,
\begin{equation}
\begin{split}
H &= N\int dx \{ A[(\partial_x \theta)^2 + \sin^2 \theta (\partial_x \phi)^2] \\
&- K \cos^2 \theta - K_{\rm IP} \sin^2 \theta \cos^2 \phi \},
\label{H2}
\end{split}
\end{equation}
where $N$ denotes the number of the spins in the cross sectional area of the model.
The third term on the right hand side is the shape anisotropy along the $x$ axis with the effective anisotropy constant $K_{\rm IP}$.
This term arises from the demagnetization between the magnetic moments, which depends on the sample shape~\cite{benjamin,skaugen}.
The effective anisotropy constant $K_{\rm IP}$ can be written as~\cite{franke2021}:
\begin{equation}
K_{\rm IP} = \frac{1}{2} \mu_0 N_x M_S^2,
\end{equation}
where $N_x$ is the demagnetizing factors,
\begin{equation}
N_x =1- [\frac{2}{\pi}\arctan (\frac{1}{p})+\frac{p}{\pi}\ln (p)+(\frac{1-p^2}{2\pi p})\ln (1+p^2)].
\end{equation} 
$p$ is the ratio between the film thickness and width.
Fig.~\ref{1}(b) shows the the width dependence of effective in-plane anisotropy $K_{\rm IP}$.
It confirms previously that a N\'eel-type domain wall transforms to a Bloch-type domain wall when the racetrack width is reduced~\cite{franke2021,dejong2015}.
For our geometry, the N\'eel-type domain walls are stable as the width of the racetrack is narrow.
Note that the demagnetization energy is written as the effective anisotropy term for analytical solutions, while it is numerically solved in simulations.

A metastable static N\'eel-type domain wall connects the anisotropy minima $\phi = 0$ or $\pi$ when $\theta = \pi / 2$ and thus satisfies the Euler-Lagrange equations
\begin{equation}
A \partial_x^2 \phi - K_{\rm IP} \sin \phi \cos \phi = 0 .
\label{euler}
\end{equation}
With the additional boundary condition $\partial_x \phi (\pm \infty) = 0$, we integrate Eq.(\ref{euler}),
\begin{equation}
A(\partial_x \phi)^2 - K_{\rm IP} \sin^2 \phi = 0.
\label{bc}
\end{equation}
Note that Eq.~(\ref{bc}) exhibits the symmetries $\phi \rightarrow -\phi$ and $\phi \rightarrow \phi + \pi$, which reflect the fact that the energy (\ref{H2}) is invariant under rotations by $\pi$.
Fig.~\ref{1}(c) shows the same normalized energy density when $\phi = 0$ and $\pi$.

\section{domain wall qubit}
Previous studies show the microscopic quantum effect by a coherent spin state path integral formalism and a collective coordinate technique~\cite{benjamin,benjamin1997}.
Here we derive the effective Hamiltonian for a domain wall qubit with the consideration of a uniform external magnetic field $\boldsymbol{B} = B \hat{y}$ along the $y$ axis.
By adding a Zeeman energy term, the total energy $H$, i.e., Eq.~(\ref{H2}) is changed into
\begin{equation}
\begin{split}
H &= N\int dx \{ A[(\partial_x \theta)^2 + \sin^2 \theta (\partial_x \phi)^2] \\
&- K \cos^2 \theta - K_{\rm IP} \sin^2 \theta \cos^2 \phi-B\sin\theta\sin\phi \}.
\label{H3}
\end{split}
\end{equation} 
The Lagrangian of the domain wall thus takes the form~\cite{Galkina,benjamin}
\begin{equation}
L=\frac{SN}{a}\int dx[\dot{\phi}(1-\cos\theta)]-H,
\label{L}
\end{equation}
where the first term is the Berry phase term, $S$ is the magnitude of the spin, and $a$ is the lattice constant.

In the presence of dissipation, the equations of motion are in polar coordinates,
\begin{equation}
\begin{split}
\dot{\theta} - \alpha \dot{\phi} \sin \theta &= \frac{2A}{\hbar} \frac{1}{\sin \theta} \nabla \cdot (\sin^2 \theta \nabla \phi) \\
&+ \frac{2K_{\rm IP} }{\hbar} \sin \theta \sin\phi\cos\phi - \gamma B \cos\theta \cos\phi,
\label{llg1}
\end{split}
\end{equation}
\begin{equation}
\begin{split}
\dot{\phi} \sin \theta + \alpha \dot{\theta} &= -\frac{2A}{\hbar}\nabla^2 \theta +\frac{2A}{\hbar}\sin \theta \cos \theta (\nabla \phi)^2 \\
&+\frac{2K}{\hbar} \sin \theta \cos \theta + \frac{2K_{\rm IP} }{\hbar}\sin\theta \cos\theta \cos^2\phi \\
&- \gamma B \sin \phi \cos \theta .
\label{llg2}
\end{split}
\end{equation}
Introducing the domain wall solution $\theta=2\arctan e^{u}$ with $u = \frac{x-X(t)}{\lambda}$ directly~\cite{Schryer} and considering their derivations, equations~(\ref{llg1}) and~(\ref{llg2}) can be rewritten as
\begin{equation}
(1+\alpha^2)\dot{\phi}=\gamma B \tan u (\sin\phi+\alpha\cos\phi)-\frac{2\alpha K_{\rm IP}}{\hbar}\sin\phi\cos\phi,
\end{equation}
\begin{equation}
(1+\alpha^2)\dot{X}/\lambda=\gamma B \tan u (\alpha \sin\phi-\cos\phi)+\frac{2K_{\rm IP}}{\hbar}\sin\phi\cos\phi.
\end{equation}
Thus, we can get the velocity $v=\dot{X}(t)$,
\begin{equation}
v=\frac{\lambda\gamma B \tan u (\alpha \sin\phi-\cos\phi)+\frac{2\lambda K_{\rm IP}}{\hbar}\sin\phi\cos\phi}{1+\alpha^2},
\label{v}
\end{equation}
and the width $\lambda$ of the moving wall
\begin{equation}
\lambda =\delta (1+\frac{K_{\rm IP}}{K}\cos^2\phi)^{-1/2},
\end{equation}
where $\delta=\sqrt{A/K}$ is the width  of the domain wall at rest.

\begin{figure}
\includegraphics{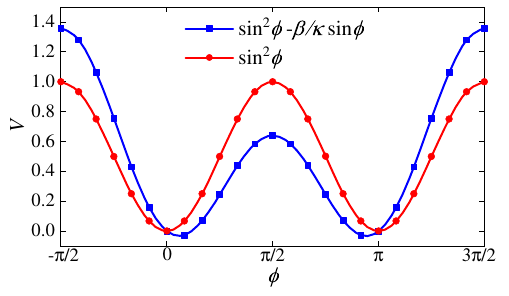}
\caption{
The potential energy as a function of $\phi$.
The red curve shows the case of zero magnetic field, while the blue curve corresponds to $B=100$ mT.
Here, $K_{\rm IP}=1.4\times10^2$ $\rm Jm^{-3}$ and $\beta/\kappa=0.2$ are adopted.}
\label{2}
\end{figure}
\begin{figure*}
\includegraphics{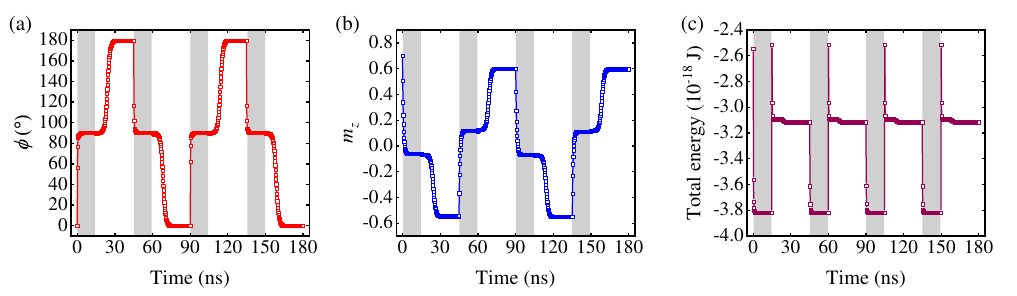}
\caption{NOT gate constructed by using the magnetic field.
(a) The reversible conversion between basic quantum bits $\lvert 0 \rangle$ and $\lvert 1 \rangle$. Angle between the center magnetization and the $x$ direction as a function of time. 
(b) The out-of-plane spin components of the domain wall as a function of time. 
(c) The total energy as a function of time. 
The magnetic field pulse duration is indicated by the gray background.}
\label{3}
\end{figure*}
If the shape anisotropy, originating from the demagnetization effect, is much smaller than the easy-axis anisotropy, the dynamics of the domain wall can be described by the action of a free particle of mass $M$~\cite{doring1948,benjamin},
\begin{equation}
M= \frac{N S^2}{a^2 K_{\rm IP}}\sqrt{\frac{K}{A}}.
\end{equation}
Therefore, the effective Lagrangian for the domain wall position $X$ is given by~\cite{shibata,benjamin}
\begin{equation}
L_X = \frac{M}{2}\dot{X}^2 + \rho C\dot{X} ,
\label{Lx}
\end{equation}
where the first term is the kinetic energy of the domain wall, and the second term has the form of a gauge potential $\rho=\pi SN/a$.
The second term depends on the chirality $C=\pm 1$ of the domain wall, which is from the Berry phase term.
Introducing the momentum of the domain wall by the canonical relation $[\hat{X}, \hat{P}]=i\hbar$, the Lagrangian~(\ref{Lx}) corresponds to the effective Hamiltonian
\begin{equation}
\hat{H}=\frac{1}{2M}(\hat{P}-\rho \sigma_z)^2,
\label{QH1}
\end{equation}
where $\sigma_z$ is the Pauli matrix characterizing the chirality $C$ of the domain wall.

Next, we consider the influence of the external magnetic field $\bm{B}$ along the $y$ axis.
Integrating out the $\theta$ fluctuations around the domain wall in Eq.~(\ref{L}), an effective Lagrangian in $\phi$ is obtained~\cite{benjamin},
\begin{equation}
L_{\phi}=\frac{M_c}{2}\dot{\phi}^2+V(\phi),
\label{Lphi}
\end{equation}
where $M_c=NS^2\pi^2\delta/8a^2K$ is the effective mass related to the chirality dynamics, and the barrier potential $V(\phi)$ is
\begin{equation}
V(\phi)=\kappa \sin^2\phi - \beta\sin\phi,
\label{potential}
\end{equation}
where $\kappa=2\delta N K_{\rm IP}$ and $\beta=g\mu_B SN\pi\delta B/a$ with the Lander factor $g$ and Boltzmann constant $\mu_B$.
The potential has two minima when $\sin\phi=\beta/2\kappa$, as shown in Fig.~\ref{2}.
The magnetic field along $y$ axis suppresses the barrier at $\phi=\pi/2$.

The level splitting $\epsilon$ is obtained by standard calculation using the instanton technique~\cite{zou2022domain},
\begin{equation}
\epsilon \approx 4\hbar\omega_0\sqrt{S_{\rm inst}/2\pi\hbar}e^{-S_{\rm inst}/\hbar},
\end{equation}
where $S_{\rm inst} \approx 4V_0/\omega_0$ is the instanton action with the tunnel barrier $V_0=2NK_{\rm IP}(1-B/2)^2$ and the level spacing $\hbar\omega_0=2\sqrt{2AK_{\rm IP}(1-B^2/4)}$.
Choosing $K_{\rm IP}=1.4\times10^2$ $\rm Jm^{-3}$ and $B=100$ mT, We find the chirality splitting $\epsilon=3.64\times10^{-22}$ $\rm J$.

Previous studies have showed that the degeneracy between the domain walls of opposite chirality is lifted when the magnetic field is applied along the propagation axis, i.e., $x$ axis~\cite{benjamin1996,benjamin1997}.
The effective Zeeman effect splits the energy between the two chirality states.
Another similar effect is achieved by applying the electric field along the easy axis, where the chirality remains unaffected by the electric fields along the $z$ axis~\cite{benjamin}.
Considering the magnetic field along the $y$ axis and the electric field along the $z$ axis, the effective Hamiltonian~(\ref{QH1}) becomes
\begin{equation}
\hat{H}=\frac{1}{2M}(\hat{P}-\rho \sigma_z)^2+\epsilon\sigma_x+H_E\sigma_z,
\label{QH2}
\end{equation}
where $H_E=E_zP_z$ is the energy term related to the electric field $E_z$ along the $z$ axis and the $z$ component of the total electric polarization $P_z$.
The electric polarization is induced by two neighboring spins under the electric fields and is defined as~\cite{katsura,xia2023}
\begin{equation}
\bm{P} = -\frac{Aea}{E_{\rm SOC}}\sum_{i,j} \bm{e}_{ij} \times (\bm{S}_i \times \bm{S}_j),
\end{equation}
where $i$ and $j$ are the sites indices, $\bm{S}_i=\bm{M}_i/M_S$ is the spin at the site $i$, $\bm{e}_{ij}$ is the unit vector connecting the two sites of neighboring $\bm{S}_i$ and $\bm{S}_j$, $e$ is the electron charge, and $E_{\rm SOC}$ is the magnitude of the spin-orbit interaction~\cite{rashba2003}.

\section{implementation of quantum gates}
With a universal set of quantum gates, any quantum algorithm can be implemented by controlling a particular unitary evolution of the qubits. 
The dynamics of the qubit state is governed by the Schr$\ddot{\rm o}$dinger equation
\begin{equation}
i\hbar\partial\lvert \psi \rangle / \partial t =\hat{H} \lvert \psi \rangle.
\end{equation}
In this part, a set of quantum gates for universal quantum computation is discussed, including the single-qubit gates (NOT gate, phase-shift gate, and Hadamard gate) and two-qubit gates (Ising gate and CNOT gate).  

\subsection{NOT gate}
The NOT gate performs the inversion of the qubit state, which is represented by the matrix $\sigma_x$,
\begin{equation}
\sigma_x = 
\begin{pmatrix}
0 & 1 \\
1 & 0 
\end{pmatrix}
.
\end{equation}
The implementation of NOT gate on spins is reduced to the rotation of the spin by $180^\circ$ by the magnetic field pulse based on the Hamiltonian~(\ref{QH2}).
Applying the magnetic field when $0\leq t \leq t_0$ and setting $\epsilon=\hbar\theta/(2t_0)$, the solution of the Schr$\ddot{\rm o}$dinger equation is
\begin{equation}
U_X (\theta) = \exp [-\frac{i\theta}{2} \sigma_x].
\label{xgate}
\end{equation}
This is the NOT gate for the angle $\theta=\pi$.
Fig.~\ref{3} depicts the process to switch between the qubit states $\lvert 0 \rangle$ and $\lvert 1 \rangle$ using magnetic field pulses on the racetrack. 
The switching time of the domain wall qubit state is 30 ns.
The initial spin configuration of the racetrack is set to state $\lvert 0 \rangle$. 
It should be noted that the magnetic field pulses remain 100 mT in the $+y$ direction throughout the process. 
By adjusting the duration and strength of the magnetic field, the NOT gate is accomplished.

\subsection{Phase-shift gate}
\begin{figure}
\includegraphics{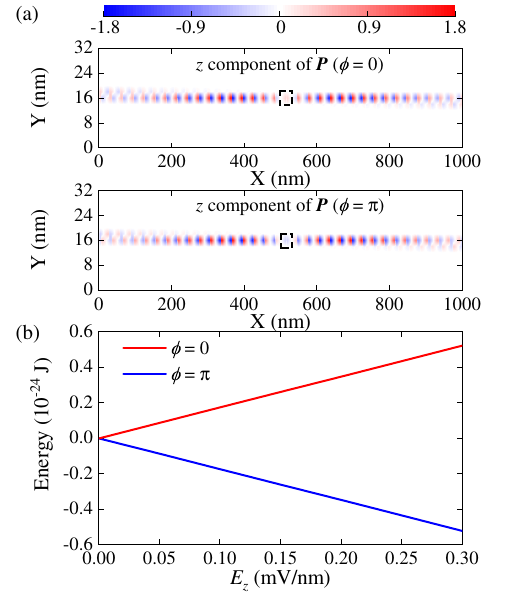}
\caption{Phase-shift gate constructed by controlling the electric field. 
(a) Spatial profile of the $z$ component of the electric polarization of domain walls with different chiralities. The color palette indicates the magnitude of the $z$ component of the electric polarization. Black dashed rectangulars show opposite polarizations of the domain walls with different chiralities. 
Following parameters are adopted: $e=1.62\times10^{-19}$ C, $a=1$ nm, and $E_{\rm SOC}=10$ meV.
(b) Electric field dependence of the energy.}
\label{4}
\end{figure}
A phase-shift gate modifies the phase of the state $\lvert 1 \rangle$ relative to the state $\lvert 0 \rangle$, and is represented by the matrix,
\begin{equation}
\rm Ph (\theta)=
\begin{pmatrix}
1 & 0\\
0 & e^{i\theta}
\end{pmatrix}
,
\end{equation}
where $\theta$ is the phase shift.
This is equivalent to a rotation around the $z$ axis by $\theta$ radians.
When $\theta=\pi$, it is also called the Pauli-Z gate.
According to the Hamiltonian~(\ref{QH2}), it can be achieved by applying electric fields to rotate the domain wall qubits.
Setting the $H_E(t)=\hbar\theta/(2t_0)$ for $0\leq t \leq t_0$ and $H_E(t)=0$ otherwise, the phase-shift gate is achieved by the solution of the Schr$\ddot{\rm o}$dinger equation
\begin{equation}
U_Z(\theta)=\exp[-\frac{i\theta}{2}\sigma_z].
\end{equation}
The phase shift $\theta$ can be changed by tuning the strength of the electric field.
Fig.~\ref{4}(a) presents the numerical spatial distribution of the $z$ component of the electric polarization, highlighting the opposite polarity between two domain walls with different chirality, as indicated by the dashed rectangles. 
The energy difference between the two chirality states increases with the applied electric field as shown in Fig.~\ref{4}(b). 
The energy difference is of the order of $10^{-24}$ J at $0.3$ mV/nm, where the operation time is 0.1 ns.

\subsection{Hadamard gate}
The Hadamard gate is defined by the matrix
\begin{equation}
U_H = \frac{1}{\sqrt{2}}
\begin{pmatrix}
1 & 1\\
1 & -1
\end{pmatrix}
\label{hadamard}
.
\end{equation}
It can be realized by a sequential application of the phase-shift gates and the NOT gate by setting $\theta = \pi/2$ as~\cite{schuch}
\begin{equation}
U_H = -i U_Z(\frac{\pi}{2}) U_X(\frac{\pi}{2}) U_Z(\frac{\pi}{2}).
\label{hadamardseries}
\end{equation}

\subsection{Ising gate}
Two-qubit gates between neighboring qubits can be performed by switching on their exchange interaction.
For two qubits, the Hamiltonian is given by
\begin{equation}
H_{\rm Ising}=J_{\rm exchange} \sigma_z^{(1)} \otimes \sigma_z^{(2)}.
\label{Hising}
\end{equation}
Setting $H_{\rm Ising}(t)=\hbar\theta/2t_0$ for $0\leq t \leq t_0$ and $H_{\rm Ising}(t)=0$ otherwise, the Ising coupling gate is obtained as
\begin{equation}
U_{ZZ} (\theta) = \exp [-\frac{i\theta}{2}\sigma_z^{(1)} \otimes \sigma_z^{(2)}],
\end{equation}
acting on the two qubits in the neighboring layers with interlayer exchange interactions.
\begin{figure}
\includegraphics{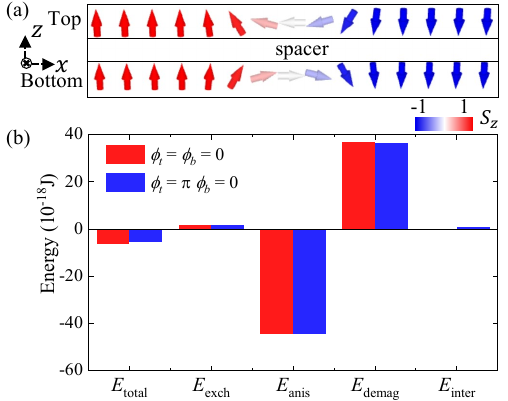}
\caption{Ising gate constructed by considering the Ising interaction between two domain walls. 
(a) The sketch of bilayer domain walls. 
(b) Energies of bilayer domain walls with the identical and opposite chiralities, including the total energy $E_{\rm total}$, the exchange energy $E_{\rm exch}$, the anisotropy energy $E_{\rm anis}$, the demag energy $E_{\rm demag}$, and the interlayer exchange energy $E_{\rm inter}$. 
The model includes two magnetic layers and a nonmagnetic spacer. 
The thicknesses of the magnetic and nonmagnetic layers are equal to 1 nm. 
$A_{\rm inter} =1.0\times10^{-15}$ $\rm J m^{-1}$ is adopt.}
\label{5}
\end{figure}
By setting $\theta=\pi$, the Ising coupling gate is expressed as matrix
\begin{equation}
U_{ZZ} \equiv 
\begin{pmatrix}
1 & 0 & 0 & 0 \\
0 & -1 & 0 & 0 \\
0 & 0 & -1 & 0 \\
0 & 0 & 0 & 1
\end{pmatrix}
.
\end{equation}
Here, we consider a bilayer racetrack, where two racetracks are placed vertically as shown in Fig.~\ref{5}a.
A domain wall with $\phi = 0$ is initially placed at the center of the bottom magnetic layer, and a domain with $\phi = \pi$ is initially placed at the center of the top magnetic layer.
A thin metallic spacer layer is introduced between two racetracks~\cite{avci,xia2022}.
We have numerically evaluated the energy difference between the identical and opposite chiralities, which is $\sim 8.7 \times 10^{-22}$ $\rm J$.
The operating time is $\sim 0.2$ ps.
The detailed energies of the bilayer domain walls are shown in Fig.~\ref{5}b.

\subsection{CNOT gate}
The CNOT gate $U_{\rm CNOT}^{1 \to 2}$ is defined by the matrix
\begin{equation}
U_{\rm CNOT}^{1 \to 2} =
\begin{pmatrix}
1 & 0 & 0 & 0 \\
0 & 1 & 0 & 0 \\
0 & 0 & 0 & 1 \\
0 & 0 & 1 & 0
\end{pmatrix}
.
\end{equation}
The CNOT gate is constructed by a sequential application of the CZ gate and the Hadamard gate as
\begin{equation}
U_{\rm CNOT}^{1 \to 2} = U_H^{(2)} U_{CZ} U_H^{(2)},
\end{equation}
where the control qubit is the domain wall in the one layer and the target qubit is the domain wall in the another layer, and the CZ gate can be realized by a series of phase-shift gates and Ising coupling gate~\cite{schuch}, i.e.,
\begin{equation}
U_{CZ} = e^{i\pi /4} U_Z^{(1)}(\frac{\pi}{2}) U_Z^{(2)}(\frac{\pi}{2}) U_{ZZ}^{(1)}(-\frac{\pi}{2}).
\end{equation}

\section{coherence time}
%
%
Coherence time quantifies the duration for which a domain wall qubit can retain its coherence before decoherence occurs. 
In the context of the domain wall qubit, the decoherence occurs because of the interactions between the qubit and its environment.
These interactions act as dissipative forces in terms of the phenomenological Gilbert damping $\alpha$ to the quantum dynamics of the domain wall qubit, resulting in relaxation and dephasing.
The relaxation time of the domain wall qubit can be mathematically described by~\cite{zou2022domain}
\begin{equation}
T^{-1} = \frac{\epsilon^2 [\hbar^2 \omega_0^2 / \rho^2 + 4 \epsilon^2+H_E^2 / (\hbar M \omega_0)]}{2(\epsilon^2+H_E^2)(\hbar^2 \omega_0 / \rho )^2}  S_{\xi}(\Omega),
\end{equation}
where $S_{\xi}(\Omega)$ is related to the dissipative parameter via the fluctuation dissipation theorem,
\begin{equation}
S_{\xi}(\Omega)= \tilde{\alpha} \hbar \Omega \coth \frac{\hbar \Omega}{2 K_B T_c}.
\end{equation}
Here, $\tilde{\alpha}=\alpha N \hbar S$, $\hbar \Omega=\sqrt{\epsilon^2+H_E^2}$, $K_B$ is the Boltzmann constant, and $T_c$ is the operational temperature.
The relaxation time of the domain wall qubit is estimated as 0.67 $\mu$s with $N=52$, $S=1$, $a=2$ nm, $\tilde{\alpha}=1.10\times10^{-36}$, $\epsilon=3.64\times10^{-22}$ $\rm J$, and $H_E=1.0\times10^{-22}$ $\rm J$, which is much longer than the quantum gate operation times and can be increased by optimizing the material parameters.
Without the application of the magnetic field along the $x$ axis, the pure dephasing is absent, resulting in a dephasing time twice of the relaxation time.

\section{initialization and readout}
For many applications in quantum computing, the initial state required is a simple pure state where all qubits are in the pure state $\lvert 0 \rangle$.
We can achieve this by applying an electric field $E_z$ to resolve the degeneracy between the left-handed and right-handed domain walls, as shown in Fig.~\ref{4}b. 
By cooling down the sample, each qubit will naturally fall into the ground state $\lvert 0 \rangle$, because the state $\lvert 0 \rangle$ has the lower energy. 
Another approach to create a domain wall with a specific chirality is by gradually cooling the system while applying a magnetic field~\cite{zou2022domain}.

Qubit readout determines the result at the end of the computation by measuring specific qubits.
Various experimental techniques can be used to determine the chirality of a domain wall, i.e., the qubit state. 
One common method is using the magnetic force microscopy~\cite{mfm}, which allows for nanoscale visualization of the structure. 
Another technique is using the Lorentz transmission electron microscopy~\cite{ltem}, which enables direct imaging with high spatial resolution. 
NV centers have also been employed as a non-invasive quantum sensor to measure the chirality of nanoscale domain walls~\cite{nv}.

\section{discussion and conclusion}
We remark the relevance to a recent theoretical report on qubits based on domain walls~\cite{zou2022domain}. 
First, although both the mechanisms use the chirality of a domain wall as a qubit, its implementation platform is different. 
The domain wall qubit, considered as one-dimensional spin chain, is manipulated in ferrimagnetic racetrack in the previous report, while we construct it in ferromagnetic racetrack with finite width, which is more realistic in applications.
Second, the DMI which prefers domain walls with certain chirality is considered as an effective magnetic field in the previous proposal, which makes it complicated when encoding the degenerate domain walls. 
However, the DMI is not necessary in our proposal, where a broad range of materials can be used.
Third, the mechanism of implementing single-qubit gate differs. 
In previous work, single-qubit gates are achieved by shaking the domain wall. 
The shaking of the domain wall may emit spin waves, which induces a complicated interaction between domain walls.
However, we employ magnetic fields and electric fields to realize single-qubit gates, offering a more manageable approach for manipulation.
Lastly, the coherence time is longer than that in the previous proposal, which is benefit for the quantum computation.

In this paper, we have proposed a theoretical approach to utilize the chirality of domain walls as a means to encode qubit states.  
Our study has focused on crucial aspects of quantum computing, including the identification of well-defined qubits, reliable state preparation, quantum gate operations, and state readout. 
The utilization of in-plane magnetic fields has been used to control the domain wall qubits for the NOT gates, electric fields for the phase-shift gates, and Ising couplings for the Ising gates.
Furthermore, we have shown that fundamental gates such as CNOT gates and Hadamard gates can be constructed through a series of phase-shift gates, NOT gates, and Ising gates. 
We also discuss the coherence time, which is estimated in the microsecond region.
These findings pave the way for the realization of efficient and scalable quantum computing architectures based on domain wall qubits.
Our results may contribute to the growing field of quantum information processing and open up new possibilities for harnessing the unique properties of domain walls in quantum technologies.

\begin{acknowledgments}
The work is supported by Shenzhen Fundamental Research Fund (Grant No. 399 JCYJ20210324120213037), Guangdong Basic and Applied Basic Research Foundation (2021B1515120047), Guangdong Special Support Project (2019BT02X030), Shenzhen Peacock Group Plan (KQTD20180413181702403), and National Natural Science Foundation of China (NSFC) (11974298).
M. E. acknowledges the support by the CREST, JST (Grants No. JPMJCR20T2) and the Grants-in-Aid for Scientific Research from MEXT KAKENHI (Grant No. 23H00171).  
\end{acknowledgments}

\bibliography{reference}

\end{document}